\documentclass[aip,jcp,preprint]{revtex4-1}

\usepackage{amsmath}
\usepackage{graphicx}

\newcommand{\Ga}{{ \Gamma}}
\newcommand{\X}{{\rm X}}
\newcommand{\Y}{{\rm Y}}
\newcommand{\M}{{\rm M}}
\newcommand{\Hbias}{{\bf H}_{\text{bias}}}
\newcommand{\Mrf}{\tilde{m}({\bf r}, f)}
\newcommand{\Mrfv}{\left|\Mrf\right|}
\newcommand{\Mrfb}{\left(\Mrf\right)}
\newcommand{\tc}[1]{\textcircled{\scriptsize #1}}

\begin{document}

\preprint{}

\title{Magnonic Band Structure, Complete Bandgap and Collective Spin Wave Excitation in Nanoscale Two--Dimensional Magnonic Crystals}

\author{D. Kumar}
\thanks{These authors have contributed equally to this work.}
\affiliation{Thematic Unit of Excellence on Nanodevice Technology, Department of Condensed Matter Physics and Material Sciences, S. N. Bose National Centre for Basic Sciences, Block JD, Sector III, Salt Lake, Kolkata 700 098, India.}

\author{J. W. K{\l}os}
\thanks{These authors have contributed equally to this work.}
\affiliation{Faculty of Physics, Adam Mickiewicz University in Poznan, Umultowska 85, Pozna\'{n}, 61-614, Poland.}

\author{M. Krawczyk}
\affiliation{Faculty of Physics, Adam Mickiewicz University in Poznan, Umultowska 85, Pozna\'{n}, 61-614, Poland.}

\author{A. Barman}
\email{abarman@bose.res.in}
\affiliation{Thematic Unit of Excellence on Nanodevice Technology, Department of Condensed Matter Physics and Material Sciences, S. N. Bose National Centre for Basic Sciences, Block JD, Sector III, Salt Lake, Kolkata 700 098, India.}

\date{\today}

\begin{abstract}
We present the observation of a complete bandgap and collective spin wave excitation in two–-dimensional magnonic crystals comprised of arrays of nanoscale antidots and nanodots, respectively. Considering that the frequencies dealt with here fall in the microwave band, these findings can be used for the development of suitable magnonic metamaterials and spin wave based signal processing. We also present the application of a numerical procedure, to compute the dispersion relations of spin waves for any high symmetry direction in the first Brillouin zone. The results obtained from this procedure has been reproduced and verified by the well established plane wave method for an antidot lattice, when magnetization dynamics at antidot boundaries is pinned. The micromagnetic simulation based method can also be used to obtain iso--frequency countours of spin waves. Iso--frequency contours are analougous of the Fermi surfaces and hence, they have the potential to radicalize our understanding of spin wave dynamics. The physical origin of bands, partial and full magnonic bandgaps has been explained by plotting the spatial distribution of spin wave energy spectral density. Although, unfettered by rigid assumptions and approximations, which afflict most analytical methods used in the study of spin wave dynamics, micromagnetic simulations tend to be computationally demanding. Thus, the observation of collective spin wave excitation in the case of nanodot arrays, which can obviate the need to perform simulations may also prove to be valuable.
\end{abstract}

\maketitle

\section{Introduction}

Analogous to photonic crystals, magnonic crystals (MCs) \cite{Nikitov2001, *Demokritov2013, *Kruglyak2010} are magnetic meta-materials designed for the propagation of spin waves (SWs).\cite{Hillebrands2002, Stancil2009, Neusser2009, *Lenk2011} Based on their design, MCs exhibit a characteristic SW dispersion relation complete with bands and, sometimes, band gaps which can be tuned by controlling material and structural parameters as well as the strength and orientation of the bias magnetic field.\cite{Mamica2012a, Yang2011} This phenomenon makes MCs useful as potential candidates for the design of SW based signal processing and logic devices.\cite{Ding2012}

The knowledge of dispersion relation of a wave propagating through a medium is necessary to understand its transmission characteristics. Although MCs have been a subject of intense study lately,\cite{Mamica2012, Au2012, Klos2012, Mandal2012, *Ciubotaru2012, *Tacchi2012, *Gubbiotti2012, *Saha2013, Venkat2013, Kumar2013} reports on a time domain numerical calculation of dispersion relations of SWs propagating in two-dimensional (2D) MCs are very rare.\cite{Kim2010,Schwarze2012,Dvornik2013} As other analytical methods are available, the use of time domain simulations and spatial Fourier transform to obtain the dispersion relation in a photonic or phononic crystal is rarely seen\cite{Jafarpour2004} as well. We hope to fill that gap in research with this work. The underlying principles, over which the procedure described here is used, has been discussed more generally by us in Ref.~\onlinecite{Kumar2012}. Here too, we essentially use a micromagnetic simulator called Object Oriented Micromagnetic Framework\cite{Donahue2002} (OOMMF) to obtain magnetization $\bf{M}$, as a function of position $\mathbf{r}$, and time $t$. Then we use a multi-domain discrete Fourier transform to obtain the desired dispersion relation: SW power as a function of wavevector ${\bf k}=(k_x, k_y)$, and frequency $f$. However, while simulating the magnetization dynamics in large (ideally infinite\cite{Kruglyak2010a}) 2D crystals, one can be expected to need far greater computational resources than during the simulations of their one-dimensional (1D) counterparts.\cite{Klos2012} Using a finite sample size may produce some spurious modes in the obtained dispersion relation.\cite{Kruglyak2010a} Thus, the use of 2D periodic boundary condition\cite{Wang2010} (PBC) becomes mandatory in order to obtain good numerical resolution in wavevector and frequency domains while consuming finite computational resources. Also, 2D crystals have more high symmetry directions when compared to their 1D analogues. Different techniques will be required to obtain the results for different directions in the 2D reciprocal space covering the entire irreducible part of the Brillouin zone (BZ).\cite{Kittel1976} Moreover, the signal which generates the waves will have to be carefully designed so that the resulting spectrum represents the physical dispersion relation of plane propagating SWs. Due to all these complications, a need to validate the results obtained here with a well established method, such as the plane wave method (PWM)\cite{Krawczyk2008} becomes very clear.

Both OOMMF and PWM solve the following Landau-Lifshitz-Gilbert (LLG) equation:\cite{Landau1935, *Gilbert2004}
\begin{eqnarray}
\frac{d {\mathbf M}}{dt}&=&-\bar{\gamma}{\bf M}\times {\bf H}_{\text{eff}}-\frac{\alpha \bar{\gamma}}{M_{\text{s}}}{{\bf M}}\times\left({\bf M}\times {\bf H}_{\text{eff}}\right)\text{.}\label{eq:llg}
\end{eqnarray}
Here, $M_{\text{s}}$ is the saturation magnetization, $\bar{\gamma}$ is the gyromagnetic ratio and $\alpha$ is the Gilbert damping constant. ${\bf H}_{\text{eff}}$ is the sum of the bias field $\Hbias$, the excitation signal ${\bf H}_{\text{sig}}$ (used in micromagnetic simulations (MS) only), the exchange field, the demagnetizing field and the magneto-crystalline anisotropy. Other factors like magnetostriction, should also be considered here if applicable.

The details of MC considered here are presented in Sec.~\ref{sec:i}. Simulation parameters and PWM  are described further in Sec.~\ref{sec:ii}.
OOMMF uses the finite difference method (FDM) to solve Eq.~\eqref{eq:llg} as an ordinary differential equation in time and space (derivatives with respect to space are hidden away in ${\bf H}_{\text{eff}}$). PWM is based on the Bloch wave formalism. As these two methods are fundamentally different in approach, some quantitative differences in results are to be expected. The results from both the methods and their differences have been discussed in Sec.~\ref{sec:results} for the antidot lattice (ADL). Due to small lattice constant, the considered system is an exchange dominated one and consequently, the differences in dispersion relations along the bias magnetic field and perpendicular to it are subtle. These differences have been explored by calculating the iso-frequency contours in the wavevector space using both MS and the PWM. The iso-frequency contours are the curves of the constant frequency plotted in the wavevector space, they are wave counterparts of the Fermi surfaces known from the theory of the solid state physics.\cite{Kittel1976} The iso-frequency contours are very important tool for the analysis of the wave propagation phenomena, giving a deep insight into direction and velocity of propagating, reflected and refracted waves in artificial crystals. Such type of analysis, while widely explored in photonic and phononic crystals for designing their metamaterials properties,\cite{Luo:201104,Enoch:2003,Li:084301} is almost absent in magnonics. Thus, developing the ability to compute these iso-frequency contours using MS can be a breakthrough in exploring magnonic metamaterials based on MC; because the MSs can be performed without approximation limited applicability of the PWM (or other analytical methods\cite{Grimsditch2004}), and thus yields experimentally realizable results even with complex magnetic configurations.

We also plot the energy spectral density and phase distributions associated with different modes in the SW spectrum in order to understand their physical origin and explain any observed partial or complete bandgaps. Finally, we use the method described here to obtain the SW dispersion relations in the case of 2D dot array where the SW propagation is mediated by inter-dot stray magnetic field as opposed to dipole-exchange interaction in ADL. This brings about an interesting change in the spectra, which is discussed in Sec.~\ref{sec:results} along with their effective properties.

\section{Method}

\subsection{\label{sec:i} Magnonic crystal lattice and material parameters}

\begin{figure}[!ht]
\includegraphics[width=\textwidth]{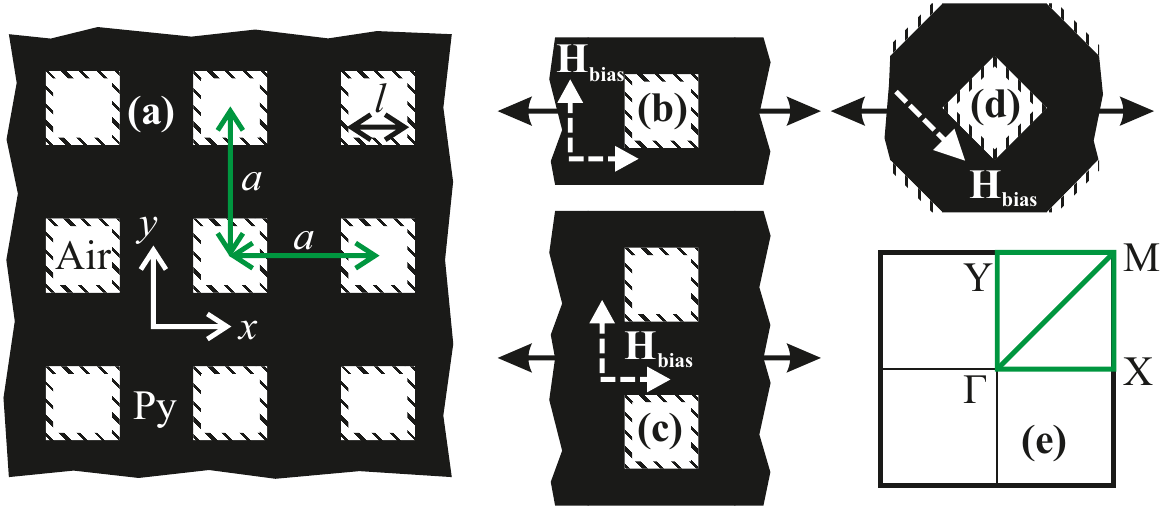}
\caption{(a) The 2D antidot lattice under consideration. The lattice constant of the square lattice is $a=30$ nm. The thickness $s$ of the film is $3$ nm. The antidots are square (white) air holes of edge $l=12$ nm in ferromagnetic Py (black) medium. Dynamics is pinned at the edge of holes. The pinned region is marked with a different texture. Element geometry used in micromagnetic simulations extends to over hundred repetitions in length (horizontal dark arrows in (b), (c) and (d)) for good wavenumber resolution. 2D PBC is applied over these elements to mimic the infinite geometry. White arrows in (b), (c) and (d) show the direction of bias field used for simulations of SW dispersion for backward volume and Damon-Eshbach configuration. (d) shows the first BZ in the inverse lattice with typical symmetry point labels.}
\label{fig:sample}
\end{figure}

The structure considered here is an infinitely large square array of square antidots with their ferromagnet-air interface under pinned boundary conditions.\cite{Klos2012} The geometrical structure of the sample is shown in Fig.~\ref{fig:sample} (a). The lattice constant $a=30$ nm and the antidots are square holes of edge length, $l=12$ nm. The material parameters  of permalloy (Py: Ni$_{80}$Fe$_{20}$) are used during simulations and in PWM calculations: exchange constant, $A=13{\times}10^{-12}$ J/m, saturation magnetization, $M_s=0.8{\times}10^6$ A/m, gyromagnetic ratio, $\bar{\gamma}=2.21{\times}10^5$ m/As and no magnetocrystalline anisotropy. A saturating bias magnetic field of $\mu_0 \Hbias = 1$ T points in $x$ direction.

\subsection{\label{sec:ii}Micromagnetic simulations and the plane wave method}

The micromagnetic simulations involve solving the LLG equation (\ref{eq:llg}) using a finite difference method based ordinary differential equation solver; and then, Fourier transforming the obtained space and time dependent magnetization data to get SW spectral density in wavevector and frequency domains.\cite{Kumar2012} 
%The results obtained from the micromagnetic simulations are validated by the plane wave method (PWM).\cite{Krawczyk2008} 
Cell size $(d, d, s) = (1.5, 1.5, 3)$ nm along ($x,y,z$) axis was used during the FDM based simulations. The pinning in micromagnetic simulations
was introduced by fixing magnetization vector in all cells of the discretization mesh, which border the antidots, i.e., in regions marked with different texture in Fig.~\ref{fig:sample}.   Figures~\ref{fig:sample} (b), (c) and (d) show parts of the elements over which 2D PBC are used to simulate the dispersion relation for different directions of the wave vector. These elements extend over $100$ (up to $300$) repetitions of unit cells in the horizontal direction to yield good resolution in the wavenumber domain. The 2D PBC are also implemented in order to improve the results with finite computational resources.\cite{Wang2010}  Figure~\ref{fig:sample} (e) shows the first BZ, the path in its irreducible part and typical symmetry points:
$\Ga = (0, 0)$, $\X = \pi/a(1, 0)$, $\Y = \pi/a(0,1)$ and $\M = \pi/a(1, 1)$.\cite{Kittel1976} Note that when the bias field is in the plane, an asymmetry is expected between the two orthogonal directions of SW propagation:  $\Hbias$$||{\bf k}$ (backward volume) and $\Hbias$${\perp}{\bf k}$ (Damon-Eshbach).\cite{Venkat2013} Thus, 
%contrary to what is mentioned in Ref.~[\onlinecite{Kim2010}], 
the triangle $\Ga \X \M$ is no longer the irreducible BZ. However, in the forward volume arrangement when $\Hbias$ is perpendicular to the plane of the 2D MC, the symmetry is restored and dispersion is the same in the two orthogonal directions. The technique described here can be used independent of the direction of $\Hbias$.

In order to get the results in the  $\Ga-\X$ and $\Y-\M$ directions, we use different excitation signals of the form ${\bf H}_{\text{sig}} = (0, 0, H_z)$, on elements shown in Figs.~\ref{fig:sample} (b) and (c), respectively. $\Hbias$ is horizontal along the $x$ axis (dashed white arrows). Similarly, dispersion along the $\Ga-\Y$ and $\X-\M$ directions can be obtained when $\Hbias$ is across the width of the elements (vertical arrows along $y$ axis). Here, $H_z=H_0N_tN(x)n_y$ with ${\mu}_0H_0=5$ mT and $N_t$, $N(x)$ and $n_y$ as given by Eqs.~\eqref{eq:Nt},~\eqref{eq:Nx} and \eqref{eq:ny}, respectively:
\begin{align}
N_t &= \frac{\sin(2{\pi}f_{\text{c}}(t-t_0))}{2{\pi}f_{\text{c}}(t-t_0)}, \label{eq:Nt} \\
N(x) &= \frac{\sin(k_{\text{c}}x)}{k_{\text{c}}x}, \label{eq:Nx} \\
n_y &= \cos(2{\pi}y/y_{\text{max}})+\sin(2{\pi}y/y_{\text{max}}). \label{eq:ny}
\end{align}
See Eq. ($3$) in Ref.~\onlinecite{Kumar2013} for the detailed description of the terms involved in these equations. Here, the origin of coordinates is at the center of the considered geometry. It is due to $N_t$ and $N(x)$ that the signal contains power between ${\pm}f_{\text{c}}$ and ${\pm}k_{\text{c}}$ in frequency and wavevector domains respectively.\cite{Kumar2012} $n_y$ should be asymmetric to ensure that both symmetric and antisymmetric modes are present in the resulting spectrum.\cite{Hillebrands2002} In Eq.~\eqref{eq:ny}, $y$ goes from $0$ to $y_{\text{max}}$. While computing dispersion along $\Ga-\X$ and $\Ga-\Y$ directions (Fig.~\ref{fig:sample} (b)), $y_{\text{max}}=a$. However, for $\Y-\M$ and $\X-\M$ directions (Fig.~\ref{fig:sample} (c)), $y_{\text{max}}=2a$. Both the elements in Figs.~\ref{fig:sample} (b) and (c) will span the same infinite 2D geometry under a 2D PBC; except, in the later case we can control whether the dynamics in the neighboring rows will be in phase or out of phase. Thus we can fix the wavevector component $k_y$ or $k_x$ to 0 or $\pi/a$ in the simulations.  This is necessary to differentiate between the parallel directions $\Ga-\X$ and $\Y-\M$ or $\Ga-\Y$ and $\X-\M$. Also, $n^{mn}_y$ given by the expression
\begin{align}
n^{mn}_y &= C_m\cos(2m{\pi}y/y_{\text{max}})+C_n\sin(2n{\pi}y/y_{\text{max}}) \label{eq:nmny} 
\end{align}
can be used instead of $n_y$ to selectively alter the amplitude of $m$-th symmetric or $n$-th antisymmetric mode. The freedom of choice of amplitudes $C_m$ and $C_n$ allows us to artificially control the statistical temperature of the magnons in the crystal and also helps in isolating a single mode in the case of a degeneracy. We can also sum over $m$ and $n$ to alter multiple modes in a single dynamic simulation.
We also attempt to obtain the dispersion in $\Ga-\M$ direction by using the element shown in Fig.~\ref{fig:sample} (d). However, as there are two scattering centers (antidots) per cell in this arrangement, we can obtain the dispersion relations correctly only up to half of the BZ in that direction.\cite{Schwarze2012} 

Until now, we could use a signal similar to the one we did in the case of an 1D lattice.\cite{Kumar2013} But, this limitation forced us to come up with a new signal
\begin{equation}
H_z=H_0N_tN(x)N(y)n(x)n(y), \label{Signal_2D}
\end{equation}
which has to be used in a larger 2D lattice of $100{\times}100$ antidot array (with  the cell size $d$ increased to $3$ nm to decrease time of computations). Here, $n(x)$ is given by:
\begin{align}
n(x) &= \sum\limits_{m=1}^{5}\left(\sin(2{\pi}mx/a)-\cos(2{\pi}mx/a)\right),\label{eq:nx}
\end{align}
with analogous formula for $n(y)$. This signal is a point like source with the amplitude decay with distance as described by sinc function (in $N(x)$ Eq.~(\ref{eq:Nx}) along $x$ axis and in similar form for $N(y)$ for $y$ dependence), having sharp cut-off in Fourier domain and able to excite symmetric and antisymmetric modes with respect to $x-$ or $y-$ axis.  This signal was arrived upon largely by intuition, nevertheless, its agreement with the results obtained from PWM validates the usefulness of this signal. Spectral density, periodicity and asymmetry of the excitation signal (or source) should also be considered while developing similar techniques for other kinds of crystals (\textit{e.g.} photonic or phononic crystals).

Three fold (one in time and two in space) Fourier transforms was needed to obtain the SW dispersion here. Magnetization was assumed to be uniform across the thickness of the film. We can now easily generalize that in the case of three-dimensional MCs, a signal of the form $H_z=H_0N_tN(x)N(y)N(z)n(x)n(y)n(z)$ will be required followed by a four fold discrete Fourier transform.

We have also calculated the spatial distribution of energy spectral density (ESD), $S_f$ and phase, $\theta$ from the following equations:
\begin{align}
S_f &= \Mrfv^2; \label{eq:sf} \\
\theta &= \tan^{-1}\left(\frac{\text{Im}\Mrfb}{\text{Re}\Mrfb}\right). \label{eq:theta}
\end{align}
Here, $\Mrf$ is the time domain Fourier transform, of a dynamical magnetization data. Unlike the new method used in Ref.~\onlinecite{Kumar2013}, this gives us power from the entire wavevector domain for a selected frequency $f$. However, if power is present for just one particular wavevector then both methods yield qualitatively identical results. 
%Spatial distribution of amplitude squared for plane propagating SWs, $\left|m_z\right|^2$ is also plotted using the PWM.

The PWM is a spectral method in which the eigenproblem is numerically solved in the frequency and wavevector domains by the standard numerical routines. We solve here LLG equation (\ref{eq:llg}) in linear approximation without damping. The PWM calculations  are performed with the assumption of the full magnetic saturation of the ADL along the bias magnetic field.  As pinning during simulation will occur at the cell's center, a hole size of $l+d$ was assumed during PWM calculations. Due to small thickness of the ADL, uniform SW profile across the thickness is assumed. The PWM in this formulation was already used in the calculations of the SW dynamics in 2D ADL and proved to give correct results.\cite{Neusser2011, Tacchi2012a, Klos2012, Klos2013} The detailed description of the method can be found in Refs.~\onlinecite{Tacchi2012a} and \onlinecite{Neusser2011a}.

\section{\label{sec:results}Results and Discussions}

\begin{figure}[!ht]
\includegraphics[width=\textwidth]{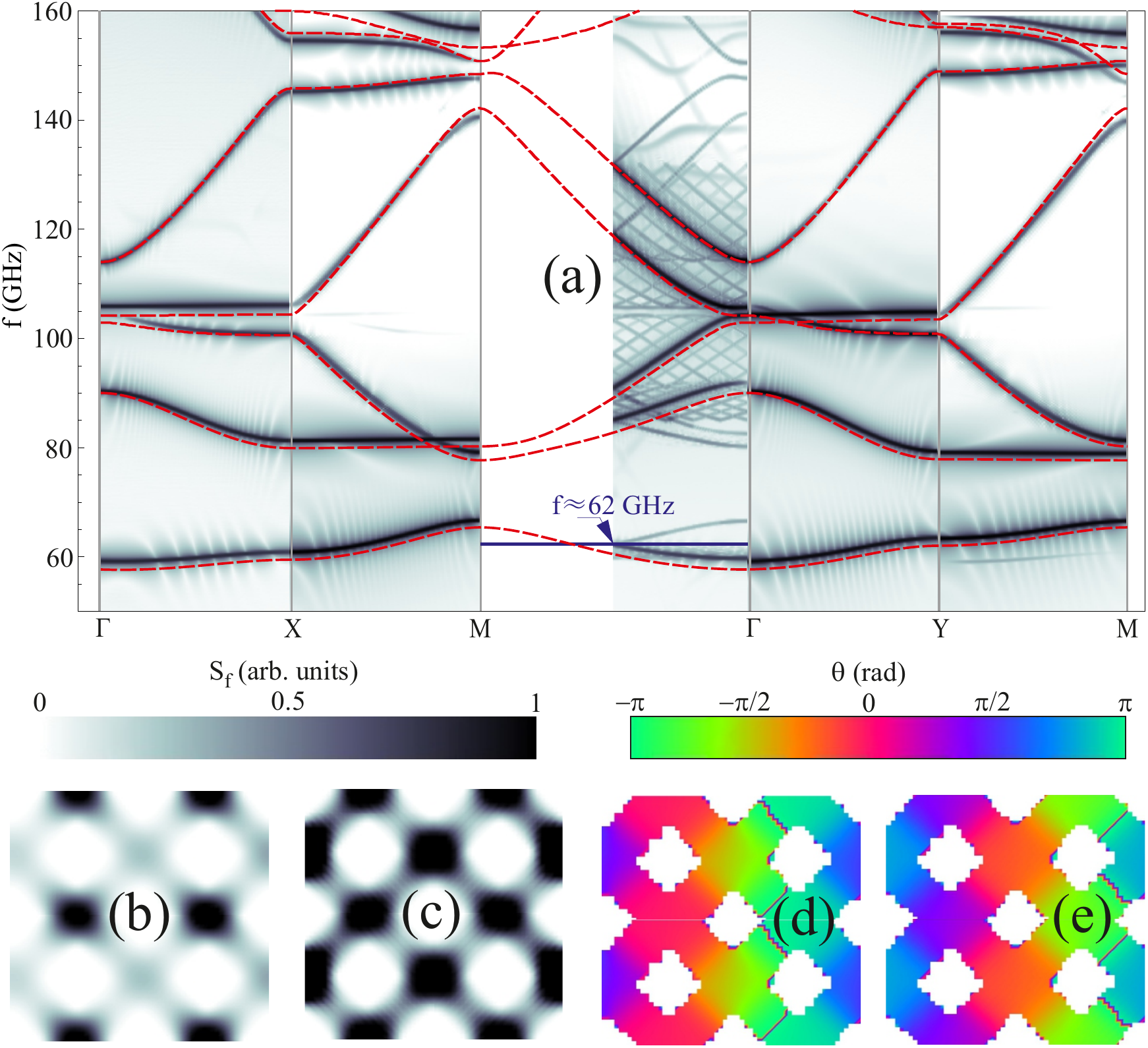}
\caption{(a) SW dispersion calculated using MSs (solid line) and PWM (dashed lines). ESD $S_f$, distribution for the horizontal line ($f \approx 62$ GHz) shown in (a) in parts of the sample when the propagation direction is along $\Ga-\M$ for (b) $k_{\text{c}} = \pi/\left(\sqrt{2}a\right)$ and (c) $k_{\text{c}} = \sqrt{2}\pi/a$. Corresponding phase $\theta$, distribution is shown in (color online) (d) and (e), respectively.}
\label{fig:disp1}
\end{figure}

The dispersion along the path in the first BZ shown in Fig.~\ref{fig:sample} (e)  calculated with MSs by using the elements shown in Fig.~\ref{fig:sample} (b)-(d) is assembled as Fig.~\ref{fig:disp1} (a) using solid lines. An overlay of dashed lines representing the SW dispersion relation obtained from the PWM is provided for comparison. Both these results appear to agree with each other except for the $\Ga-\M$ direction where the numerical method was able to yield results for only half of the total BZ extent. This is because we set $k_{\text{c}}$ to $\pi/\left(\sqrt{2}a\right)$ here (the spatial periodicity is $\sqrt{2}a$). Compared to the element shown in Fig.~\ref{fig:sample} (b), which can be used to produce results for $\Ga-\X$ or $\Ga-\Y$ directions, the one in Fig.~\ref{fig:sample} (d) features two scatter centers per unit cell. And, if we artificially increase $k_{\text{c}}$ to $\sqrt{2}\pi/a$, both scattering centers will be activated to produce additional spurious modes.\cite{Schwarze2012} To demonstrate the same we plot $S_f$ (normalized between $0$ and $1$) and $\theta$ (given by Eqs.~\eqref{eq:sf} and \eqref{eq:theta}, respectively), for frequency $f \approx 62$ GHz  in Figs.~\ref{fig:disp1} (b) to (e). Note that the horizontal separation between regions of high ESD is about $\sqrt{2}a$ in Fig.~\ref{fig:disp1} (b) for $k_{\text{c}} = \pi/\left(\sqrt{2}a\right)$. This reduces to $a/\sqrt{2}$ in Fig.~\ref{fig:disp1} (c) for $k_c = \sqrt{2}\pi/a$ when both scattering centers in the unit cell (of the element shown in Fig.~\ref{fig:sample} (d)) are activated at once. The phase distributions also confirm that neighbouring locations of high ESD are about $\pi$ and $\pi/2$ radians out of phase with each other in former ($k_{\text{c}} = \pi/\left(\sqrt{2}a\right)$: Fig.~\ref{fig:disp1} (d)) and later ($k_{\text{c}} = \sqrt{2}\pi/a$: Fig.~\ref{fig:disp1} (e)) cases, respectively. Apart from incomplete result for the $\Ga-\M$ direction, we can also see that the modes here (shown by solid lines) do not match with those for $\Ga-\Y$ direction at the $\Ga$ point. This is because (cell size) $d = \sqrt{2}$ nm was used while simulating for the $\Ga-\M$ direction as opposed to $d = 1.5$ nm, which was used in the case of $\Ga-\Y$ direction. Also, there are additional modes of lower amplitudes visible in the case of $\Ga-\M$ direction. This is due to the fact that $N(x)$ becomes a stepped approximation of the right hand side of \eqref{eq:Nx} by the use of the FDM; thus compromising the effectiveness of the cut off at $k_{\text{c}} = \pi/\left(\sqrt{2}a\right)$, and exciting the second scattering centre to some extent (but not as well as $k_{\text{c}} = \sqrt{2}\pi/a$).

\begin{figure}[!ht]
\includegraphics[width=15 cm]{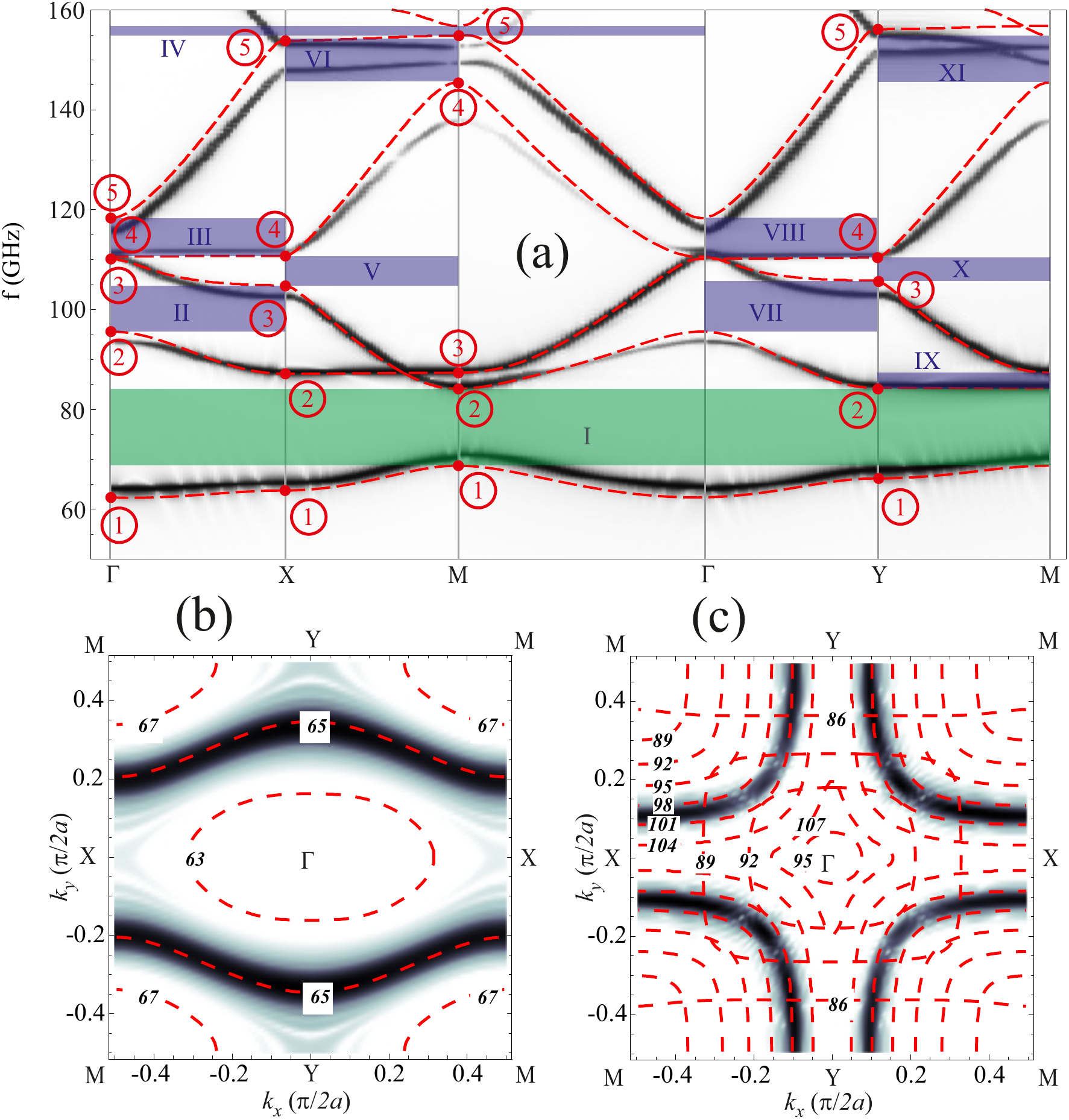}
\caption{(a) SW dispersion calculated using MS (solid line) and the PWM (dashed lines). The full and partial magnonic bandgaps are marked and numbered by Roman numerals. The circled Arabic numerals indicate the points on the dispersion for which the mode profiles are calculated in Fig.~\ref{fig:modes}. Iso-frequency lines from (b) $63$ GHz to $67$ GHz (c) $86$ GHz to $107$ GHz (it is around the top and bottom of the first and second magnonic band, respectively) using the PWM is shown with dashed lines. Iso-frequency lines for (b) $f \approx 67$ GHz and (c) $f \approx 100$ GHz calculated by the numerical method is shown using solid lines.}
\label{fig:disp2}
\end{figure}

In pursuit of our quest to close the gap in the $\Ga-\M$ direction we eventually decided to simulate the SW dynamics in a large 2D MC with signal defined by Eq.~(\ref{Signal_2D}) and perform a three-fold Fourier transform in contrast with the two-fold transforms done earlier. We transformed time to frequency domain and $x-$ and $x-$ dimensions to the 2D wavevector domain. The resulting dispersion relation as calculated from numerical method is shown in Fig.~\ref{fig:disp2} (a) using solid lines. The agreement with the PWM results (shown by dashed lines) is poorer in comparison with Fig.~\ref{fig:disp1} (a). This is due to the fact that cell size in the later attempt was increased from $d = 1.5$ nm to $d = 3$ nm. The complete and partial bandgaps width and center frequency, as seen from the dashed lines in Fig.~\ref{fig:disp2} (a), are extracted in Tab.~\ref{tab:bg}. Here, values for partial bandgaps depend upon the path, which has been used to plot the dispersion. Bandgap I is the only complete bandgap observed here with the maximum width of $15.37$ GHz.
%If $\Ga-X$-M was to be mistakenly treated as the irreducible BZ,\cite{Kim2010} then bandgap IV would also have erroneously registered as a full bandgap. 

\begin{table}[!ht]
\caption{\label{tab:bg}Magnonic bandgap widths and center frequencies across different directions as calculated by the PWM and labeled in Fig.~\ref{fig:disp2} (a).}
\begin{tabular}{|c|c|c|c|}
\hline Label & Extent & Center (GHz) & Gap Width \\ 
       &  &  & (GHz) \\ 
\hline I & Complete Bandgap & 76.39 & 15.36 \\ 
\hline II & $\Ga-\X$ & 100.18 & 9.24 \\ 
\hline III & $\Ga-\X$ & 114.45 & 7.7 \\ 
\hline IV & $\Ga-\X-\M-\Ga$ & 155.85 & 1.9 \\ 
\hline V & $\X-\M$ & 107.75 & 5.9 \\ 
\hline VI & $\X-\M$ & 149.6 & 8.4 \\
\hline VII & $\Ga-\Y$ & 100.68 & 10.24 \\
\hline VIII & $\Ga-\Y$ & 114.2 & 8.2 \\ 
\hline IX & $\Ga-\Y$ & 85.80 & 3.01 \\
\hline X & $\Y-\M$ & 108.1 & 4.6 \\ 
\hline XI & $\Y-\M$ & 150.1 & 9.4 \\
\hline
\end{tabular}
\end{table}

Most bands observed in Fig.~\ref{fig:disp2} (a) increase or decrease almost monotonously along any high symmetry direction. Consequently, the width of bandgap I too, appears to decrease monotonously as we move either along $\Ga \rightarrow \X \rightarrow \M$ or $\Ga \rightarrow \Y \rightarrow \M$. Both upper and lower limits of bandgap I are present at point $\M$ which suggests an anti-crossing of bands at that point. This can also be regarded as the cause of the gap formation. Narrower bandgap widths have been observed by different techniques before.\cite{Wang2010a} The relatively high width of $15.37$ GHz of bandgap I here can be attributed to small lattice dimensions and edge pinning.\cite{Klos2012} Bandgaps II to XI (Fig.~\ref{fig:disp2} (a) and Tab.~\ref{tab:bg}) are direction dependent partial bandgaps. This is mainly because bands approaching point $\M$ from other high symmetry directions (with the exception of the band starting at $\Ga$\tc{5}) tend to show greater slopes. As $\X$\tc{5}$ \rightarrow \Y$\tc{5} is a relatively flatter line, bandgap IV survives for three high symmetry directions. In a more isotropic forward volume arrangement,\cite{Schwarze2012} bandgap IV might also have qualified as a complete bandgap if the dispersion in the $\X - \M$ direction was also calculated. On the other hand, if wavevector dependent anisotropy is overlooked,\cite{Kim2010} partial bandgaps (e.g. bandgap IV, or II and VII, or III and VIII) will appear as a complete bandgap. Partial bandgaps IV, V, VIII, X and XI are direct, while II, III, VI, VII and IX are indirect. Direct bandgaps are formed when the minimal  and  the maximal frequency of the magnonic bands surrounded the bandgap, from the top and bottom, respectively, are characterized by the same wavevector. While two different wavevectors are involved in the formation of indirect bandgap. In Fig.~\ref{fig:disp2} (a) the minimal and maximal frequencies appear at high symmetry points. Occasionally, a bandgap may form between two high symmetry points due to anti-crossing of modes in a folded BZ,\cite{Klos2013} but that is not observed here.

\begin{figure}[!ht]
\includegraphics[width=0.48 \textwidth]{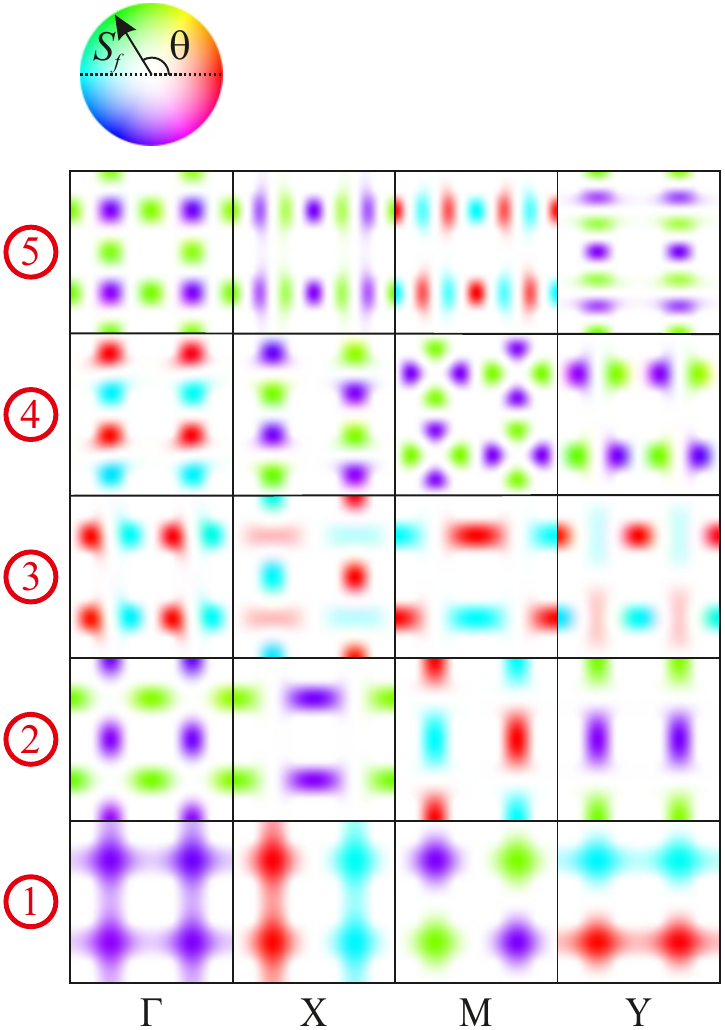}
\caption{(Color online) ESD $S_f$, and phase $\theta$, for high symmetry points $\Ga$, $\X$, $\M$ and $\Y$ at points \tc{1} through \tc{5} marked on Fig.~\ref{fig:disp2} (a).}
\label{fig:modes}
\end{figure}

Now we calculate mode profiles ESD $S_f$ and phase $\theta$, at the high symmetry points, using the PWM,  for the first five modes as marked in Fig.~\ref{fig:disp2} (a). The results are tabulated as Fig.~\ref{fig:modes} where $S_f$ is represented by color saturation and $\theta$ is represented by hue. A general trend of higher frequency mode profiles limiting themselves to smaller regions in real space is observed. This trend has been seen for 1D systems as well.\cite{Klos2013} Here, mode profiles appear similar in size at points $\X$\tc{5}, $\M$\tc{5} and $Y$\tc{5}. Although, the distribution at $\Y$\tc{5} is vastly different due to a (nearby) mode--crossing in the $\Y-\M$ direction (see Fig.~\ref{fig:disp2} (a)). Mode profile at $\Y$ may be obtained by rotating the mode profiles at $\X$ by $90^\circ$. Modes with negligible group velocity are trapped and forbidden to move in specific high symmetry directions. Also, the number of nodal lines, which controls the spatial quantization of modes, generally increases with mode number \tc{i}$: \text{i}\in \left\lbrace 1, 2, 3, 4, 5\right\rbrace$. No nodal lines are evident for $\Ga$\tc{1}. Vertical and horizontal nodal lines are seen at $\X$\tc{1} and $\Y$\tc{1}, respectively; while $\M$\tc{1} features both vertical and horizontal nodal lines. From Fig.~\ref{fig:disp2} (a), we can see that points $\Ga$\tc{2}, $\X$\tc{2}, $\M$\tc{3} and $\Y$\tc{3} belong to the same mode and points $\Ga$\tc{3}, $\X$\tc{3}, $\M$\tc{2} and $\Y$\tc{2} belong to a different mode. As the crossing between these modes occurs along the $\X-\M$ direction, the mode profiles at $\X$\tc{2} and $\M$\tc{3} are comparable. Similarly, mode profiles at $\X$\tc{3} and $\M$\tc{2} are also comparable except, $\X$\tc{3} has higher frequency and consequently, is more confined is space. In general, vertical and horizontal nodal lines dominate at points $\X$ and $\Y$, respectively; while a more isotropic distribution is observed at point $\Ga$ and $\M$. Modes \tc{1}, \tc{2}, and \tc{5} are isotropic along $x-$ and $y-$axes for the $\Ga$ point. However, modes \tc{3} and \tc{4} are disposed along rows and columns, respectively. There local shape and size is comparable and accordingly, they are also degenerate as seen in Fig.~\ref{fig:disp2} (a). Going from $\Ga$ to either $\X$ or $\Y$, \tc{4} maintains its size and frequency; except the Damon-Eshbach\cite{Leite2001} geometry is evident in the later case. Similarly, the expanses of mode profiles at \M\tc{2} and \tc{3} are comparable (as there frequencies are within $5$ GHz of each other), and yet their orientations are mutually orthogonal.

Iso-frequency lines are shown in Figs.~\ref{fig:disp2} (b) and (c), using both the PWM (dashed lines) and the MSs (solid lines). Iso-frequency contours calculated using the proposed method are thicker because small $a$ yields a low wavevector resolution. The agreement between the results obtained from the two methods as $67$ GHz line calculated using the MSs and the $65$ GHz line calculated using the PWM is clear, but the 2 GHz difference in frequencies is due to the shift of the dispersion curves calculated with both methods shown in Fig.~\ref{fig:disp2} (a). In contrast to Fig.~\ref{fig:disp2} (b) the two methods appear to give identical results for the $100$ GHz iso-frequency line, where the results of MS and PWM coincide. The shapes of iso-frequency lines control the direction of the propagating waves and consequently also alter the shapes of their wavefronts. Thus, although the dispersion along $\Ga-\X$ and $\Ga-\Y$ directions may appear comparable, the wavefronts of the propagating SWs from the first band will quickly uncover the underlying anisotropy, because of slightly different group velocity and curvature of different iso-frequency contours in two orthogonal directions, which is easily noticeable by the inspection of the contours for 63 and 65 GHz in Fig.~\ref{fig:disp2} (b). This anisotropy is a manifestation of dipolar interactions hardly visible in this size and frequency regime in the magnonic band structure shown in Fig.~\ref{fig:disp2} (a).
Backward volume modes are characterized by negative group velocity in the case of dipole dominated or dipolar-exchange SW propagating in a ferromagnetic thin film.\cite{Stancil2009} This is not seen in Fig.~\ref{fig:disp2} (a) as due to weakness of the dipolar interactions the exchange field makes a significant contribution with increasing wavevector ${\bf k}$ already near the BZ center.

\begin{figure}[!ht]
\includegraphics[width=0.48 \textwidth]{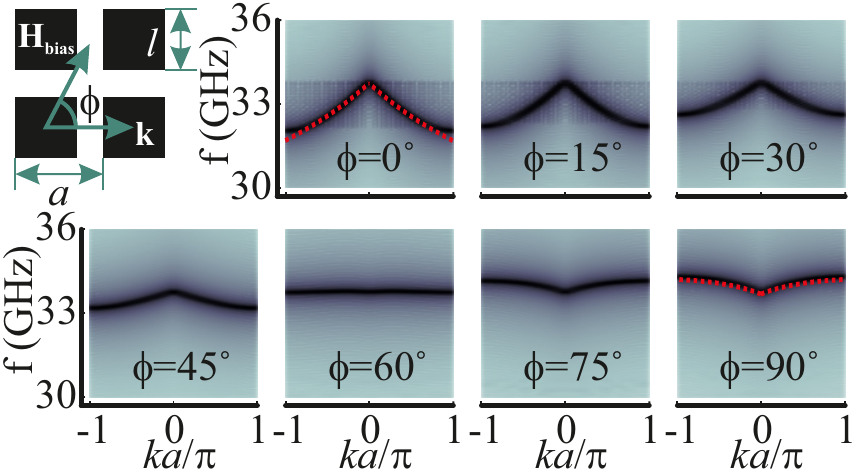}
\caption{First mode in a permalloy nano--dot array with varying angle $\phi$, between the bias field $\Hbias$ ($\mu_0$$\Hbias$ = 1 T), and wavevector ${\bf k}$, showing the transition from magnetostatic backward volume mode to Damon-Eshbach configuration. The dashed lines are calculated using the analytic expressions for these two configuration with a reduced saturation magnetization. The structure considered here is given in the top left corner with $a=9$ nm, $l=6$ nm and thickness $s = 3$ nm. Material parameters remain the same as before.}
\label{fig:dot}
\end{figure}

The developed method is not limited to the antidot lattices nor exchange dominated SWs. To prove this and better understand the properties of dipolar waves in MCs we take a look at the dispersion of SWs in the case of 2D MC composed of a square array (of lattice constant $a=9$ nm) of square dots (of edge $l=6$ nm and $3$ nm thick). This structure is shown in the top left panel of Fig.~\ref{fig:dot} along with the dispersion relations of the first mode with increasing angle $\phi$ (from $\phi = 0$ to $\phi = 90^\circ$), between $\Hbias$\ and ${\bf k}$ in the subsequent panels. Here the wave propagation is mediated by the dipolar field only. We note how the mode's group velocity gradually increases from negative (backward volume) to positive (Damon-Eshbach) as $\phi$ goes from $0^\circ$ to $90^\circ$.\cite{Dvornik2013} The transition appears to occur at a critical angle $\phi = \phi_{\text c} \approx \pi/3$. Note that here the direction of $\Hbias$\ is being changed as opposed to that of ${\bf k}$ in the previous case. It is interesting to note that the dispersion relations obtained here for the array of nano--dots reminds us of the dispersion of magnetostatic waves in thin ferromagnetic film. To verify this hypothesis we calculate the dispersion relation of magnetostatic waves in the thin ferromagnetic magnetic film (3 nm thick) with reduced magnetization, \textit{i.e.} with the effective value of the saturation magnetization $M_{\text{s,eff}}$. The dashed lines overlaid in Fig.~\ref{fig:dot} are computed using the analytical expression for backward volume and Damon-Eshbach configuration in the case of thin film\cite{Stancil2009} with a reduced saturation magnetization $M_{\text{s,eff}} = M_{\text{s}} l^2 /a^2$. A good agreement between the dispersion in the array and the effective thin film is found. A minor disagreement is introduced by the presence of the BZ boundaries but only near these boundaries. Further, the critical angle, $\phi_{\text c} =\tan^{-1} \sqrt{H_{\text{bias}} /M_{\text{s,eff}}}$~\cite{Hurben1995} in the case of such thin film is also $56.24^\circ \approx \pi/3$. This implies that one should also be able to use the analytical expression to calculate the SW manifold between backward volume and Damon-Eshbach geometries. This also shows, that a thin film MC composed of an array of saturated ferromagnetic nanodots can be used as a magnonic metamaterial, i.e., an artificial crystal with tailored effective properties of spin wave dynamics.\cite{KruglyakIntech,Zivieri2012e127,Mruczkiewicz12,Dmytriiev.104405} Further studies are necessary to elucidate the limits of the effective saturation magnetization approach presented here, the influence of the dot--shape, their arrangement and inter--dot separation (mode-splitting has been experimentally demonstrated for nano--dot arrays\cite{Rana2011}), but this is outside the scope of this paper.

\section{Conclusions}

We have described a numerical algorithm to calculate the dispersion of plane propagating SWs in a 2D MC using multi-domain Fourier transform of results obtained from micromagnetic simulations. At the core of this technique is a new excitation signal, which is capable of generating SWs whose energy spectral density corresponds to the characteristic dispersion relation of the 2D MC. The lack of such signal has been discussed before in the case of 1D MCs.\cite{Di2013, *Lee2013} The results obtained from this procedure were verified by the plane wave method when magnetization dynamics at antidot boundaries is pinned. We noted that both methods were in qualitative agreement with each other. The fact that better quantitative agreement was observed while using 2D PBC over 1D elements was due to lower cell size.\cite{Klos2013}

Apart from a new numerical algorithm to compute the dispersion relation in any given direction of a two- or three-dimensional inverse lattice, this method will also allow for the numerical computation of iso-frequency contours from micromagnetic simulations. Thus the numerical tool for study metamaterials properties of MCs was provided. It gives the possibility for design the properties of SWs relevant to technological applications and potentially exceeding these known from the homogeneous ferromagnetic thin films. The negative refraction, unidirectional media or caustic propagations are only some of examples here.\cite{Veerakumar.214401,Kostylev.134101,Kim:212501} Further, this method can be generalized to aid the numerical computation of dispersion or iso-frequency contours in the case of two- or three-dimensional phononic\cite{Alleyne1991} and photonic\cite{Burt2004, *Gersen2005} crystals as well.

The dispersion here appeared to be similar in $\Ga-\X$ and $\Ga-\Y$ directions. However, a noticeable anisotropy between the backward volume and Damon-Eshbach geometries was very evident from the study of the mode profiles and the iso-frequency contours. As dipole field mediates the SW propagation in a 2D dot array we were able to obtain the negative group velocity associated with the first mode in the case of a backward volume magnetostatic configuration. We were also able to analyze the nature of bands and complete and partial bandgaps that were obtained from the dispersion calculations in the case of an MC. This can be useful in the design of attenuators,\cite{Sekiguchi2012} phase-shifters,\cite{Au2012} filters\cite{Kim2009} and logic gates.\cite{Khitun2008}

Recent advances in lithography techniques\cite{Mandal2012, Rahman2012, Sidorkin2009, Siegfried2011} have made it possible to fabricate dot and antidot lattices with a resolution below 10 nm. Thus, one can fabricate samples with dimensions comparable to the systems considered here. Experimental techniques similar to Brillouin light scattering spectroscopy\cite{Perzlmaier2005} can be used to explore the SW dispersion relation.

\begin{acknowledgments}
We acknowledge the financial support from the Department of Science and Technology, Government of India (Grant nos. INT/EC/CMS (24/233552), Department of Information Technology, Government of India (Grant no. 1(7)/2010/M\&C), the European Community's FP7/2007-2013 (GA nos. 233552 (DYNAMAG) and  228673  (MAGNONICS)) and NCN of Poland (DEC-2-12/07/E/ST3/00538).  D.K. would like to acknowledge financial support from CSIR - Senior Research Fellowship (File ID: 09/575/(0090)/2011 EMR-I) and fruitful discussions with A. Mookerjee at the S. N. Bose National Centre for Basic Sciences.
\end{acknowledgments}

%\bibliographystyle{aipnum4-1}
%\bibliography{RefDB}

%merlin.mbs aipnum4-1.bst 2010-07-25 4.21a (PWD, AO, DPC) hacked
%Control: key (0)
%Control: author (8) initials jnrlst
%Control: editor formatted (1) identically to author
%Control: production of article title (-1) disabled
%Control: page (0) single
%Control: year (1) truncated
%Control: production of eprint (0) enabled
%

\end{document}